# Resonance order–dependent plasmon-induced transparency in orthogonally-arranged nanoscale cavities

## Naoki Ichiji,[1] Atsushi Kubo[2,*]

[1]*Graduate School of Pure and Applied Sciences, University of Tsukuba, 1-1-1 Tennodai, Tsukuba-shi, Ibaraki 305-8573, Japan*
[2]*Faculty of Pure and Applied Sciences, University of Tsukuba, 1-1-1 Tennodai, Tsukuba-shi, Ibaraki 305-8573, Japan*
*\*kubo.atsushi.ka@u.tsukuba.ac.jp*

**Abstract:** In this study, we investigate plasmon-induced transparency (PIT) in a resonator structure consisting of two orthogonally-arranged metal-insulator-metal (MIM) nanocavities with the aim of spectral modulation of a specific resonant order of the resonator. Our FDTD simulations demonstrate that when both cavities in this structure resonate at the same frequency, the PIT effect can be used to induce spectral modulation. This spectral modulation depends on the resonance order of the cavity coupled directly to the external field, occurring when first-order resonance is exhibited, but not with second-order resonance. We confirmed that this behavior is caused by the discrepancies between odd-order and even-order resonances using classical mechanical models analogous to the nanocavities. By tuning the resonance frequency and resonance order of the cavities, one can modulate the spectrum of the resonator structure in an order-selective manner.

The creation of metamaterials composed of engineered nanoscale resonator structures (also known as meta-atoms) has led to the realization of significant advances in optical manipulation [1]. The control of optical properties enabled by metamaterials is one of the foundations for controlling light in free space, and surface plasmon polaritons (SPPs) [2–4]. The unique properties of metamaterials are attributed to the dispersion relations near the resonant frequencies of the meta-atoms. The development of modulation techniques and explanation of the resonant spectrum of meta-atoms are fundamental topics in the study of metamaterials. The resonant wavelengths of many meta-atoms, such as Fabry-Perot type resonators, are due to the reflection of free electrons or SPPs at their physical boundaries. Resonance spectra typically exhibit multiple peaks, originating from multiple resonance orders [5]. These peaks are observed when the incident wave has a spectral width that spans multiple orders of resonance [6]. However, as it is difficult to suppress or modulate resonances of a specific order exclusively, the working band of a metamaterial is limited to the resolution of the constituent meta-atoms.

The electromagnetically-induced transparency effect is a quantum phenomenon that arises from the interference between different excitation pathways in a three-level atomic system. An EIT-like effect known as plasmon-induced transparency (PIT) [7–9], can be used to modulate the spectrum of a plasmonic resonator. As a plasmonic analog of EIT, PIT occurs due to the interference between plasmonic structures with electric or magnetic resonances, such as metal rods and cavities [10–12]. The coupling between the resonator excited directly by external light (the bright resonator) and the indirectly-excited resonator (the dark resonator) generates significant modulation in the resonance spectrum [13–15]. Owing to its large spectral modulation effect and high sensitivity, the PIT phenomenon has been applied in many fields, such as slow-light generation [16], sensing [17], and waveguide development [18].

In this study, we investigate the resonance interaction between the two metal-insulator-metal (MIM) cavities in an orthogonally-arranged resonator structure to realize spectral modulation depending on the resonance order. Only one of the cavities in this structure is directly excited by SPP wave packets (WP), which are injected next to the resonator structure on the metal surface. This cavity thus functions as a bright resonator directly coupled to the external field, while the other cavity functions as a dark resonator that is excited indirectly via the bright resonator. An FDTD simulation revealed that the spectral modulation due to the PIT phenomenon, which occurs when the two cavities have similar resonance wavelengths, depends on the resonance order of the bright resonator. The mode-splitting characteristic of PIT was observed when the bright resonator exhibited first-order resonance, while the modulation of spectral shape was dependent on the resonance wavelength of the dark resonator. Conversely, no mode-splitting was observed when the bright resonator exhibited a second-order resonance, and the effect of the dark resonator on the resonance spectrum was marginal. This resonance order–dependent PIT phenomenon was confirmed through analysis of classical mechanical models analogous to the nanocavities.

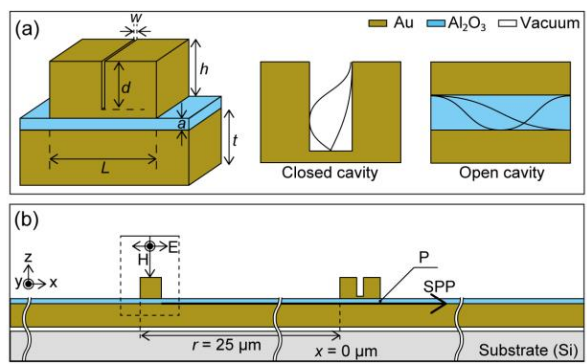

Fig. 1. Schematic of the multilayered structure used in FDTD simulation. (a) Illustration of the orthogonally-arranged resonator and individual cavities. (b) Illustration of the resonator in the simulation area.

Figure 1(a) shows a schematic of the resonator structure studied in this work as defined in FDTD simulation. A commercial FDTD software (FDTD Solution, Lumerical, Inc.) was used for all simulations. Here, we placed an Au block ($h$ =100 nm) on an $Al_2O_3$ (thickness $a$: 16 nm)/Au (thickness $t$: 100 nm) layer to form an open-ended horizontal MIM cavity (hereafter called the "open cavity") [19,20]. The resonant modes of this open cavity are given by the following equations [6,21,22]:

$$Lk_0 n_{OC} + \phi_{OC} = N\pi, \qquad (1)$$

where $k_0$ is the vacuum wave number, $n_{OC}$ is the real component of the effective refractive index of the MIM nanocavity, $N$ is the integer defining the order of the resonant mode, and $\phi_{OC}$ is the additional phase shift resulting from the opening edge.

We defined the second cavity by including a narrow slit with width $w$ and depth $d$ in the Au block. Thus, the Au block has the same geometry as a U-shaped resonator, which is a typical magnetic resonator structure [23–25]. When the slit width of this structure is sufficiently thin, it can be regarded as a half-closed MIM nanocavity (hereafter termed the "closed cavity"). The resonant mode for the closed cavity is given by the following equation [21,26–28]:

$$dk_0 n_{CC} + \phi_{CC} = \left(N - \frac{1}{2}\right)\pi, \qquad (2)$$

where $n_{CC}$ is the real component of the effective refractive index of the closed MIM nanocavity and $\phi_{CC}$ is the additional phase shift resulting from the opening edge. From Eqs. (1) and (2), the resonance frequency of each cavity is connected either to the cavity length, $L$, or the slit depth, $d$. Therefore, the resonance wavelength of the two cavities constituting the resonator can be defined independently.

We placed an Au ridge structure 25 μm from the resonator (Fig. 1 (b)), to act as an SPP excitation source. SPP WPs with a broad spectral range were excited by injecting the ridge with ultrashort 1.2 fs pulses with a peak wavelength of 600 nm. These WPs propagate on the metal surface, entering the open cavity from the side. In this configuration, the open cavity can be regarded as the bright resonator, as it is directly excited by the SPP WP, while the closed cavity can be regarded as the dark resonator, as it is excited indirectly via the resonance of the open cavity.

To determine the resonator's spectral response, we prepared a reference model without meta-atoms that was otherwise identical in construction to our initial model. The time evolution of the vertical component of the electric field ($E_z(t)$) was subsequently recorded at point $P$, located at the right edge of the resonator structure. Then, we evaluated $R(\omega)$, the spectral response as,

$$R(\omega) = \frac{|F_{res}(\omega)|^2}{|F_{Ref}(\omega)|^2}, \qquad (3)$$

where $F_{res}(\omega)$ and $F_{ref}(\omega)$ are the fast Fourier transform (FFT) of the $\boldsymbol{E_z(t)}$ waveform recorded at point $P$ in the resonator model and reference model, respectively.

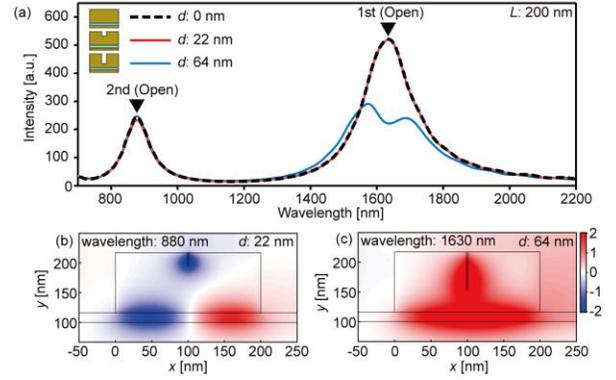

Fig. 2. (a) Spectral responses of orthogonally-arranged resonators ($L$= 200 nm and $w$= 2 nm) with differing slit depths, $d$. (b) Magnetic field ($H_y$) distribution in a resonator with a slit depth of 22 nm at a wavelength of 810 nm. In this condition, the open cavity exhibits second-order resonance, while the closed cavity exhibits first-order resonance. (c) Magnetic field distribution in a resonator with a slit depth of 64 nm at a wavelength 1630 nm. In this condition, both cavities exhibit first-order resonance.

Figure 2 (a) depicts the spectra obtained for resonator structures with $L$ fixed to 200 nm and $w$ fixed to 2 nm. Peaks can be observed at 1630 nm and 880 nm in the spectrum for a resonator with no slit in the open cavity ($d$ = 0 nm, black dashed line), corresponding to its 1st and 2nd Fabry-Perot resonant modes, respectively. As the resonance wavelength of the closed cavity is determined by slit depth, this parameter can be adjusted such that both cavities in the structure are resonant at the same wavelength. With $d$ = 64 nm, both cavities exhibit first-order resonances at 1630 nm. Because of the resulting mode-coupling, splitting can be observed in the spectrum for this structure (solid blue line). In contrast, although both cavities resonate at 880 nm with $d$ = 22 nm, here, the closed cavity exhibits first-order resonance, while the open cavity exhibits second-order resonance. Hence, no modulation is observed in the spectrum for this structure (solid red line), which is similar in shape to the spectrum of the open cavity without a slit.

The magnetic distributions shown in Figs. 2 (b) and (c) indicate that the line symmetry of the magnetic field of the open cavity with $x = L/2$ as an axis depends on resonance order. The magnetic field of the open cavity is antisymmetric with respect to the closed cavity when the former cavity exhibits even-order resonance. In such cases (e.g., for second-order resonance), the magnetic field in the region of the closed cavity originating from the open cavity is canceled out, and the closed cavity is excited predominantly by diffraction from the Au block. In contrast, when the open cavity exhibits odd-order resonance, such as first-order resonance, the induced magnetic field excites the closed cavity. The mutual inductance of the magnetic fields of the open and closed cavities results in mode-coupling [10].

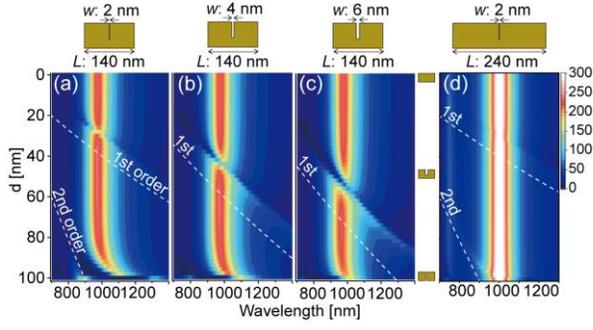

Fig. 3. Variation of resonance spectrum with slit depth for a resonator with $L$ = 140 nm and slit widths of (a) 2 nm, (b) 4 nm, and (c) 6 nm. In this condition, the open-cavity exhibits first-order resonance at 1000 nm. (d) Variation of resonant spectrum with slit depth for a resonator with $L$ = 240 nm and a slit width of 2 nm. In this condition, the open-cavity exhibits a second-order resonance at a wavelength of 1000 nm. The white-dashed lines in these images indicate the predicted resonance wavelength of the closed cavity at the corresponding slit depth, while the colors indicate the intensity of the spectral response.

Figures 3 (a), (b), and (c) show colormaps of the spectrum of resonators with slit widths of 2 nm, 4 nm, and 6 nm, respectively. The color in these images indicates the spectral response of the resonators, while the slit depth is given by the vertical axis. Here, $L$ was fixed at 140 nm, such that the open cavity exhibited first-order resonance at a wavelength of 1000 nm. The white dashed lines in each graph indicate the theoretical value of the resonance wavelength of the closed cavity, obtained by setting $\varphi_{CC} = 0$ in Eq. (2), according to the dispersion calculation for an MIM waveguide [6, 22]. The three graphs show large modulations in the resonance spectrum along the predicted resonance wavelength curves, with the slope of this modulation depending on the slit width. The shift of the predicted resonance curves in the $y$-direction is explained by the additional phase shift, $\varphi_{CC}$, in Eq. (2), which dictates the slit depth at which modulation is maximized. Spectral splitting due to the mode-coupling of resonators with the same resonance wavelength is a typical feature of PIT, while the shift of each split peak with the resonance wavelength of the dark resonator is also consistent with previous studies [29,30]. In contrast, the resonance spectrum for the structure where $L$ = 240 nm, which exhibited second-order resonance at 1000 nm, was minimally affected by the slit depth of the closed cavity, as shown in Fig. 3(d).

The dependence of the PIT phenomenon on the resonance order of the open cavity does not depend on the resonance order of the closed cavity. In Fig. 2 (a), we can confirm the large PIT-induced mode splitting along the second-order resonance curve at around $d$ = 100 nm, whereas there is no mode splitting in Fig. 2 (d) at the same $d$. The resonant order of the open cavity affects the line symmetry of the magnetic field with $x = L/2$ as an axis. However, the resonant order of the closed cavity does not affect the symmetry in the $x$-direction. Therefore, the PIT phenomenon in this configuration of the resonator structure depends only on the resonant order of the open cavity.

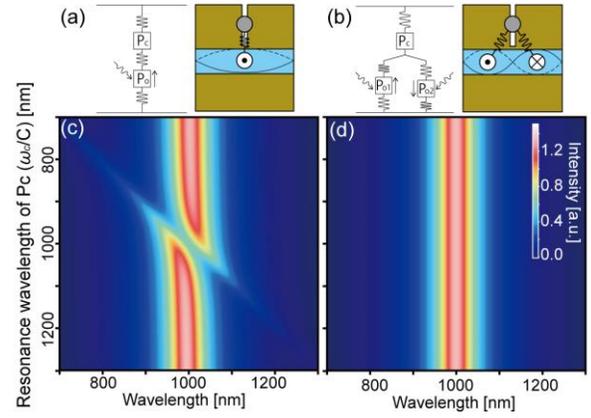

Fig. 4. (a) Schematic of the first-order system of linearly-coupled Lorentzian oscillators, modelling the case when both cavities exhibit first-order resonance. (b) Schematic of the second-order system of linearly-coupled Lorentzian oscillators, modelling the case when the open cavity exhibits second-order resonance. Variation of susceptibility with the resonance wavelength of the dipole corresponding to the closed cavity ($P_c$) calculated using the (c) first-order model, and (d) second-order model.

To verify the observed dependence of spectral modulation on resonance order, we constructed numerical models based on a classical mechanical analog to PIT. First, we defined a system consisting of two magnetic dipoles, each representing the open and closed cavities ($Po$ and $Pc$, respectively), as a model for the case when both exhibit first-order resonance ("first-order model"), as shown in Fig. 4 (a). The dynamic equation of this is described by a that of a system of linearly-coupled Lorentzian oscillators as follows [7,13,31]:

$$\begin{pmatrix} P_C \\ P_O \end{pmatrix} = \begin{pmatrix} \omega_C^2 - \omega^2 - i\gamma_C\omega & -\Omega^2 \\ -\Omega^2 & \omega_O^2 - \omega^2 - i\gamma_O\omega \end{pmatrix}^{-1} \times \begin{pmatrix} g_c E_0 \\ g_o E_0 \end{pmatrix}, \quad (4)$$

where $c$ and $o$ indicate the closed and open cavity, respectively, $\omega$ indicates a resonant wavelength, $\gamma$ indicates the damping constant for a dipole, and $g$ indicates a constant for the coupling between a dipole and the incident field. $\Omega$ indicates the constant for the coupling between the two dipoles.

In addition, we constructed a second-order model consisting of a single dipole representing the closed cavity ($P_C$), and two dipoles representing the open cavity ($P_{O_1}$, $P_{O_2}$), as shown in Fig. 4(b). Here, $P_{O_1}$ and $P_{O_2}$ represent the two peaks of the magnetic distribution when the open cavity exhibits second-order resonance, as shown in Fig. 2 (b). The dynamic equations for this model can be defined in a similar form to Eq. (4), as follows [32,33]:

$$\begin{pmatrix} P_C \\ P_{O_1} \\ P_{O_2} \end{pmatrix} = \begin{pmatrix} \omega_C^2 - \omega^2 - i\gamma_C\omega & -\Omega^2 & -\Omega^2 \\ -\Omega^2 & \omega_O^2 - \omega^2 - i\gamma_O\omega & 0 \\ -\Omega^2 & 0 & \omega_O^2 - \omega^2 - i\gamma_O\omega \end{pmatrix}^{-1}$$
$$\times \begin{pmatrix} g_c E_0 \\ g_{o1} E_0 \\ g_{o2} E_0 \end{pmatrix}. \quad (5)$$

Here, we set $g_{O_2}$, the coupling constant between $P_{O_2}$ and the incident field, equal to $-g_{O_1}$, to represent the $\pi$-phase shift between the two magnetic oscillations observed with second-order resonance. To simplify calculation, we assume that the closed cavity is not coupled to the external field at all. Hence, $g_c = 0$. In practice, this coupling constant would have a finite value because of diffraction over the top of the Au block. The values of the remaining parameters were defined as follows: $\omega_O = \omega_{O_1} = \omega_{O_2} = 300\,\text{THz}$, $\gamma_c = 25, \gamma_O = 12.5$, $\Omega = 75$, $g_{O_1} = -g_{O_2} = 1$, and $E_0 = 1$.

The susceptibility of the dipole response can be obtained from Eq. (4) and (5), $\chi_O = P_O/E_0$ and $\chi_{O_1} = P_{O_1}/E_0$. Figures 4(c) and (d) show color maps of the imaginary parts of $\chi_O$ and $\chi_{O_1}$, representing the energy dissipation of the model. Here, the resonance wavelength of the open cavity, $\lambda_o = 2\pi c/\omega_o$, was fixed at 1000 nm, while the vertical axis corresponds to the resonance wavelength of the closed cavity ($\lambda_c = 2\pi c/\omega_c$) as determined by slit depth. As with the FDTD simulations, mode-splitting was observed with the first-order model when the resonance frequencies of the two dipoles coincided ($\lambda_c = \lambda_o$), and the peaks shifted according to $\lambda_c$. In contrast, uniform spectra were observed with the second-order model regardless of the value of $\omega_c$, as shown in Fig. 4(d). These calculation results are consistent with the FDTD simulation results shown in Fig. 2 and Fig. 3.

In conclusion, we investigated the spectral modulation of a resonator consisting of an orthogonal arrangement of two types of MIM nanocavity. FDTD simulations revealed mode-splitting in the resonance spectra due to the PIT phenomenon, and peak-shifts according to the resonance frequency of the cavity acting as the dark resonator. The PIT phenomenon was highly dependent on the resonance order of the cavity acting as a bright resonator. The dependence of mode-splitting and peak shifts on resonance order was confirmed using a classical mechanical model. The shape of the resonance spectrum of the resonator structure used in this study is determined by the geometry of the cavities, the materials used for each cavity (through the effective refractive index), the alignment of the resonator with the external field, and the distance between the cavities. This demonstrates that resonance spectra design, including narrowing of line-widths, shifting of peak positions, and generation of multiple peaks, can be enabled by fabrication of composite meta-atoms consisting of bright and dark resonators. By changing the number and position of the slits, there is a possibility to modulate a specific resonance order. For experimental demonstration, it is necessary to design the structure on a sub-wavelength scale or a two-dimensional arrangement, which is suitable for the fabrication process. In addition, the dependence of mode-coupling on the mismatch between even and odd resonance order offers the potential for modulation of a specific resonant order of a meta-atom, which are the building blocks of metamaterials for broadband incident waves such as femtosecond pulses and supercontinuum waves.

**Funding.** Japan Society for the Promotion of Science, KAKENHI (JP18967972, JP20J21825); Japan Science and Technology Agency, CREST (JPMJCR14F1); Ministry of Education Culture, Sports, Science and Technology, Quantum Leap ATTO (JPMXS0118068681)

**Acknowledgments.** The authors thank H. T. Miyazaki for advice and valuable discussions.

**Disclosures.** The authors declare no conflicts of interest.

**Data availability.** Data underlying the results presented in this paper are not publicly available at this time but may be obtained from the authors upon reasonable request.

**References**

1. N. Meinzer, W. L. Barnes, and I. R. Hooper, "Plasmonic meta-atoms and metasurfaces," Nat. Photonics **8**, 889–898 (2014).
2. N. Yu and F. Capasso, "Flat optics with designer metasurfaces," Nat. Mater. **13**, 139–150 (2014).
3. Y. Liu, S. Palomba, Y. Park, T. Zentgraf, X. B. Yin, and X. Zhang, "Compact Magnetic Antennas for Directional Excitation of Surface Plasmons," Nano Lett. **12**, 4853–4858 (2012).
4. P. Genevet, D. Wintz, A. Ambrosio, A. She, R. Blanchard, and F. Capasso, "Controlled steering of Cherenkov surface plasmon wakes with a one-dimensional metamaterial," Nat. Nanotechnol. **10**, 804–809 (2015).
5. N. Ismail, C. C. Kores, D. Geskus, and M. Pollnau, "Fabry-Perot resonator: spectral line shapes, generic and related Airy distributions, linewidths, finesses, and performance at low or frequency-dependent reflectivity," Opt. Express **24**, 16366–16389 (2016).
6. N. Ichiji, Y. Otake, and A. Kubo, "Spectral and temporal modulations of femtosecond SPP wave packets induced by resonant transmission/reflection interactions with metal-insulator-metal nanocavities," Opt Express **27**, 22582–22601 (2019).
7. S. Zhang, D. A. Genov, Y. Wang, M. Liu, and X. Zhang, "Plasmon-induced transparency in metamaterials," Phys. Rev. Lett. **101**, 047401 (2008).
8. R. Taubert, M. Hentschel, J. Kastel, and H. Giessen, "Classical Analog of Electromagnetically Induced Absorption in Plasmonics," Nano Lett. **12**, 1367–1371 (2012).
9. N. Liu, L. Langguth, T. Weiss, J. Kastel, M. Fleischhauer, T. Pfau, and H. Giessen, "Plasmonic analogue of electromagnetically induced transparency at the Drude damping limit," Nat. Mater. **8**, 758–762 (2009).
10. P. C. Wu, W. T. Chen, K. Y. Yang, C. T. Hsiao, G. Sun, A. Q. Liu, N. I. Zheludev, and D. P. Tsai, "Magnetic plasmon induced transparency in three-dimensional metamolecules," Nanophotonics **1**, 131–138 (2012).
11. R. Yahiaoui, J. A. Burrow, S. M. Mekonen, A. Sarangan, J. Mathews, I. Agha, and T. A. Searles, "Electromagnetically induced transparency control in terahertz metasurfaces based on bright-bright mode coupling," Phys. Rev. B **97**, 155403 (2018).
12. M. Wan, Y. Song, L. Zhang, and F. Zhou, "Broadband plasmon-induced transparency in terahertz metamaterials via constructive interference of electric and magnetic couplings," Opt. Express **23**, 27361–27368 (2015).
13. X. Hu, S. Yuan, A. Armghan, Y. Liu, Z. Jiao, H. Lv, C. Zeng, Y. Huang, Q. Huang, Y. Wang, and J. Xia, "Plasmon induced transparency and absorption in bright-bright mode coupling metamaterials: a radiating two-oscillator model analysis," J. Phys. D: Appl. Phys. **50**, 025301 (2017).
14. Y. Lian, G. Ren, H. Liu, Y. Gao, B. Zhu, B. Wu, and S. Jian, "Dual-band near-infrared plasmonic perfect absorber assisted by strong coupling between bright-dark nanoresonators," Opt. Commun. **380**, 267–272 (2016).
15. Z. Ye, S. Zhang, Y. Wang, Y. Park, T. Zentgraf, G. Bartal, X. Yin, and X. Zhang, "Mapping the near-field dynamics in

plasmon-induced transparency," Phys. Rev. B **86**, 155148 (2012).
16. B. Zhang, H. Li, H. Xu, M. Zhao, C. Xiong, C. Liu, and K. Wu, "Absorption and slow-light analysis based on tunable plasmon-induced transparency in patterned graphene metamaterial," Opt. Express **27**, 3598–3608 (2019).
17. N. Liu, T. Weiss, M. Mesch, L. Langguth, U. Eigenthaler, M. Hirscher, C. Sonnichsen, and H. Giessen, "Planar Metamaterial Analogue of Electromagnetically Induced Transparency for Plasmonic Sensing," Nano Lett. **10**, 1103–1107 (2010).
18. C. Xiong, H. Li, H. Xu, M. Zhao, B. Zhang, C. Liu, and K. Wu, "Coupling effects in single-mode and multimode resonator-coupled system," Opt. Express **27**, 17718–17728 (2019).
19. F. Ding, Y. Yang, R. A. Deshpande, and S. I. Bozhevolnyi, "A review of gap-surface plasmon metasurfaces: fundamentals and applications," Nanophotonics **7**, 1129–1156 (2018).
20. A. Pors and S. I. Bozhevolnyi, "Plasmonic metasurfaces for efficient phase control in reflection," Opt. Express **21**, 27438–27451 (2013).
21. H. T. Miyazaki and Y. Kurokawa, "Controlled plasmon resonance in closed metal/insulator/metal nanocavities," Appl. Phys. Lett. **89**, 211126 (2006).
22. Y. Kurokawa and H. T. Miyazaki, "Metal-insulator-metal plasmon nanocavities: Analysis of optical properties," Phys. Rev. B **75**, 035411 (2007).
23. W. L. Hsu, P. C. Wu, J. W. Chen, T. Y. Chen, B. H. Cheng, W. T. Chen, Y. W. Huang, C. Y. Liao, G. Sun, and D. P. Tsai, "Vertical split-ring resonator based anomalous beam steering with high extinction ratio," Scientific Reports **5**, 11226 (2015).
24. B. Liu, C. J. Tang, J. Chen, N. Xie, J. Yuan, H. Tang, and X. Zhu, "Metal-substrate-enhanced magnetic dipole resonance in metamaterials for high-performance refractive index sensing," Optical Materials Express **8**, 2008–2016 (2018).
25. J. Chen, S. Qi, X. Hong, P. Gu, R. Wei, C. Tang, Y. Huang, and C. Zhao, "Highly sensitive 3D metamaterial sensor based on diffraction coupling of magnetic plasmon resonances," Results Phys. **15**, 102791 (2019).
26. B. Choi, M. Iwanaga, Y. Sugimoto, K. Sakoda, and H. T. Miyazaki, "Selective Plasmonic Enhancement of Electric- and Magnetic-Dipole Radiations of Er Ions," Nano Lett. **16**, 5191–5196 (2016).
27. L. Emeric, C. Deeb, F. Pardo, and J. L. Pelouard, "Critical coupling and extreme confinement in nanogap antennas," Opt. Lett. **44**, 4761–4764 (2019).
28. S. J. Park, Y. B. Kim, Y. J. Moon, J. W. Cho, and S. K. Kim, "Tuning of polarized room-temperature thermal radiation based on nanogap plasmon resonance," Opt. Express **28**, 15472–15481 (2020).
29. L. Guan, J. Zhang, Y. Xu, J.. Zhang, and Y. Li, "Impact of the Coulomb Potential and the Electrostatic Potential on the Eigen-Frequencies of the Coupled Plasmons," Plasmonics **15**, 351–359 (2020).
30. J. X. Zhang, J. Zhang, and Y. F. Li, "Plasmons Coupling and Anti-crossing of Nanometal Asymmetric Dimer," Plasmonics, (2021).
31. C. L. G. Alzar, M. A. G. Martinez, and P. Nussenzveig, "Classical analog of electromagnetically induced transparency," Am. J. Phys. **70**, 37–41 (2002).
32. K. Zhang, C. Wang, L. Qin, R. W. Peng, D. H. Xu, X. Xiong, and M. Wang, "Dual-mode electromagnetically induced transparency and slow light in a terahertz metamaterial," Opt. Lett. **39**, 3539–3542 (2014).
33. J. A. Souza, L. Cabral, R. R. Oliveira, and C. J. Villas-Boas, "Electromagnetically-induced-transparency-related phenomena and their mechanical analogs," Phys. Rev. A **92**, 023818 (2015).